\documentclass[usegraphicx,usenatbib]{mn2e}
\newcommand{\fpp}{fundamental plane}
\newcommand{\re}{R_e}
\newcommand{\Md}{M_{dyn}}
\newcommand{\sige}{\sigma_e}

\title[Mergers and the Fundamental Plane]{
Dissipationless Mergers of Elliptical Galaxies
and the Evolution of the Fundamental Plane}

\author[M. Boylan--Kolchin, C.--P. Ma, and E. Quataert]
{Michael~Boylan--Kolchin$^{1}$\thanks{E-mail:
    mrbk@astro.berkeley.edu}, Chung--Pei~Ma$^2$\thanks{E-mail:
    cpma@astro.berkeley.edu}, and Eliot~Quataert$^2$\thanks{E-mail:
    eliot@astro.berkeley.edu} \\ 
  $^1$Department of Physics, University of California, Berkeley, CA
  94720, USA\\
  $^2$Department of Astronomy, University of California, Berkeley, CA,
  94720, USA}
\begin{document}

\pagerange{\pageref{firstpage}--\pageref{lastpage}} \pubyear{2005}

\maketitle

\label{firstpage}

\begin{abstract}
  We carry out numerical simulations of dissipationless major mergers
  of elliptical galaxies using initial galaxy models that consist of a
  dark matter halo and a stellar bulge with properties consistent with
  the observed fundamental plane.  By varying the density profile of
  the dark matter halo (standard NFW versus adiabatically contracted
  NFW), the global stellar to dark matter mass ratio, and the orbit of
  the merging galaxies, we are able to assess the impact of each of
  these factors on the structure of the merger remnant.  Our results
  indicate that the properties of the remnant bulge depend primarily
  on the angular momentum and energy of the orbit; for a
  cosmologically motivated orbit, the effective radius and velocity
  dispersion of the remnant bulge remain approximately on the
  fundamental plane.  This indicates that the observed properties of
  elliptical galaxies are consistent with significant growth via late
  dissipationless mergers.  We also find that the dark matter fraction
  within the effective radius of our remnants increases after the
  merger, consistent with the hypothesis that the tilt of the
  fundamental plane from the virial theorem is due to a varying dark
  matter fraction as a function of galaxy mass.

\end{abstract}


\section{Introduction} 
\label{sec:intro}
Galaxy merging is thought to be the dominant process in the formation
and evolution of elliptical galaxies (e.g., \citealt{toomre}).  This
picture is broadly supported by both observations and simulations and
fits naturally into the $\Lambda$CDM hierarchical cosmology (e.g.,
\citealt{sn02}).  Much, however, remains to be understood about the
gas and stellar dynamics during mergers.  For example, recent
simulations indicate that not all mergers of gas-rich spiral galaxies
result in a bulge-dominated remnant \citep{sh05}.

Although gas-rich starbursts may form many of the stars contained in
elliptical galaxies, it is also likely that stars are later assembled
into larger remnants via nearly dissipationless mergers, particularly
during the formation of groups and clusters \citep{gao04}.  There is
in fact direct observational evidence for ``red'' mergers (lacking
significant star formation), both in local observations with the Sloan
Digital Sky Survey (SDSS; Masjedi and Hogg, private communication) and
in a luminous X-ray cluster at $z=0.83$ \citep{vd99}.  In addition,
the semi-analytic study of \citet{kb03} found that the most recent
merger of a luminous elliptical galaxy ($M_B \la -21$) is likely to
have been between two bulge-dominated galaxies rather than two
gas-rich spirals, again pointing to the importance of dissipationless
merging for the evolution of elliptical galaxies.

Observationally, early type galaxies exhibit a well-defined
correlation among their effective radii $\re$, luminosities $L$ (or
equivalently surface brightness $I \propto L/\re^2$), and central
velocity dispersions $\sigma$.  This ``fundamental plane''
\citep{dd87, dressler87} is often expressed as
\begin{equation}
  \re \propto \sigma^a I^b
\label{eqn:fp}
\end{equation}
and has been measured by many groups with varying results.  For
example, \citet{jorgensen96} measured $a=1.24 \pm 0.07$ and $b=-0.82
\pm 0.02$ from a set of 225 early-type galaxies in nearby clusters
observed in the r-band.  This is consistent with the original
observations by \citet{dd87} and \citet{dressler87} but noticeably
different from the SDSS determination (based on $\sim9000$ early-type
galaxies) of $a=1.49 \pm 0.05$ and $b=-0.75 \pm 0.01$
\citep{bernardi3}, values more similar to the K-band fundamental plane
of \citet{pahre98a}.  The reason for this discrepancy is not currently
clear.

Projections of the fundamental plane are also of interest in studies
of galaxy evolution.
The SDSS team measured a radius-luminosity relation of 
\begin{equation}
  \re \propto L^{0.63 \pm 0.025}
\end{equation}
 and a Faber-Jackson (1976) relation of
\begin{equation}
  \sigma \propto L^{0.25 \pm 0.012}
\end{equation}
\citep{bernardi2}, while work by \citet{kauffmann03} on the stellar
mass content of SDSS galaxies allowed \citet{shen03} to determine the
scaling of effective radius with stellar mass:
\begin{equation}
  \re \propto M_*^{0.56} \,.
\end{equation}

The \fpp\ has typically been interpreted as a manifestation of the
virial theorem, which relates $\sigma$ and $R$ to the {\it total}
enclosed (dynamical) mass $\Md$ and predicts a correlation in the
fundamental properties of galaxies:
\begin{equation}
  \sigma^2 \propto \frac{\Md}{R} = \left(\frac{\Md}{L} \right) 
  \left(\frac{L}{R^2} \right) R \propto \left(\frac{\Md}{L}\right)IR \,,
\end{equation}
giving the relation
\begin{equation}
  R \propto \sigma^2 I^{-1} \left(\frac{\Md}{L} \right)^{-1}.
\end{equation}
Clearly this virial theorem expectation is incompatible with the
observed \fpp\ in equation~(\ref{eqn:fp}) if light traces the
dynamical mass ($\Md \propto L$).  A systematic variation of $\Md/L$
with luminosity is usually assumed to ``tilt'' the virial theorem
scalings into the observed ones.  Since
\begin{equation}
  {\Md \over L} = \left( {\Md \over M_*}\right) \left({ M_* \over L}\right)\,,
\label{MoverL}
\end{equation}
this tilt could result from increasing $M_*/L$ with $L$ due to, for
instance, varying metallicity or stellar population age, or from
increasing $\Md/M_*$ with $L$ due to higher dark matter fraction in
the central parts of more luminous galaxies.  Conflicting conclusions
have been reported in the literature as to which of these explanations
is correct (for instance, compare \citealt{gerhard01} with
\citealt{pad04}).  Structural or dynamical non-homology could also
contribute to the \fpp\ tilt (e.g., \citealt{ccc95,gc97,pahre98b}).

Numerical simulations are a powerful way to assess the dynamics of
galaxy mergers.  There is a substantial body of work devoted to
simulations of mergers of galaxies in the context of forming
ellipticals (for reviews, see \citealt{bh92} and \citealt{bn03}).
Several studies have in particular addressed the question of whether
collisionless mergers of galaxies preserve the fundamental plane
relations.  These studies typically used simulations of $\sim 10^4$
particles (per initial galaxy) and have reached varying and sometimes
conflicting results.  For instance, \citet{ggva} studied mergers of
one-component systems (stellar bulges without dark matter haloes) with
$\re \propto M^{0.5}$ and found that the merger remnants ended up very
near the \fpp\ defined by the progenitors.  \citet{dantas03} found
that mergers of bulge-only galaxies remained on the fundamental plane,
while mergers of galaxies containing both bulge and dark matter
components produced a fundamental plane with a tilt larger than that
observed.  \citet{nipoti03} used repeated mergers of one- and
two-component galaxies to follow the merging process over a large
range in mass.  They found that neither the $\re-M_*$ nor the
$M_*-\sige$ relation was preserved in their dissipationless mergers
but the full fundamental plane of their merger remnants agreed with
observations.

In this paper we carry out dissipationless mergers of early type
galaxies, focusing on major mergers of two-component (dark matter plus
stellar bulge) systems.  
We thus do not address the origin of either bulge-dominated galaxies
or the fundamental plane but rather investigate whether
dissipationless mergers maintain the tight fundamental plane
relations, a scenario that is likely important for massive
ellipticals.  We pay particular attention to using cosmologically
motivated models that include adiabatic contraction of dark matter
haloes and merger orbits drawn from cosmological simulations.  We also
use a significantly larger number of simulation particles than many
previous studies, which is necessary to avoid spurious two-body
relaxation (and an accompanying artificial expansion of the stellar
bulge).  In the next section, we describe our initial models and
choice of simulation parameters in more detail.  Section
\ref{sec:properties} discusses properties of both the stellar and dark
matter components of the merger remnants in our simulations, including
their radial density profiles, triaxiality, if and how the fundamental
plane scalings and structural homology are preserved, and the changes
in the mass ratio $M_{dyn}/M_*$ in the inner parts of the remnants.
Section \ref{section:discuss} includes a discussion and summary of our
results.

\section{Simulations}

\subsection{Galaxy Models and Initial Conditions}

Our model galaxies consist of two components, a stellar bulge and a
dark matter halo.  We model the initial spatial distribution of the
bulge with a \citet{hern90} density profile,
\begin{equation}
  \rho_*(r)=\frac{M_*}{2 \pi a^3} \frac{1}{r/a} \left(\frac{1}{1+r/a}
      \right)^3 
\label{hern}
\end{equation}
with total stellar mass $M_*$, quarter-mass radius $r_{1/4}=a$, and
half-mass radius $r_{1/2}=(1+\sqrt{2})a$.  Observationally, a useful
radius is the effective radius $\re$, defined as the radius of the
isophote that encloses half of the total (projected) stellar
luminosity.  For equation~(\ref{hern}), the effective radius is
related to the quarter- and half-mass radii by $\re= 1.8153 a=0.752
r_{1/2}$.  The Hernquist profile in projection follows the
\citet{dv48} $r^{1/4}$ law to within 35\% in the radial range $0.06
\le R/\re \le 14.5$ \citep{hern90}, a region containing over 90\% of
the model's total luminosity.

For the dark matter component, we consider two possible radial density
profiles: the NFW profile \citep{nfw97} and a more realistic version
of this profile that uses the adiabatic contraction approximation
\citep{blum86} to model the response of the dark matter halo to
baryons.  The uncompressed NFW profile is given by
\begin{equation}
  \rho_{dm}(r) = \frac{\rho_c \delta_c} {(cr/r_v)(1+cr/r_v)^2} \,,
\label{nfw}
\end{equation}
where the concentration parameter $c$ is defined as the ratio of the
virial radius $r_v$ to the radius at which the logarithmic slope of
the density profile equals $-2$ (also known as the scale radius
$r_s$).  The additional parameter $\delta_c$ depends on $c$ and the
choice of virial overdensity $\Delta_v$ relative to the critical
density $\rho_c$:
\begin{equation}
  \delta_c= \frac{\Delta_v c^3}{3 [\ln(1+c)-c/(1+c)]} \,.
\end{equation}
We use $\Delta_v=200$ throughout this paper.  Since
equation~(\ref{nfw}) is logarithmically divergent as $r \rightarrow
\infty$, we use the exponential truncation scheme described by
\citet{sw99} for our haloes at $r > r_v$.  Truncating the haloes in this
manner requires $\approx 10\%$ more simulation particles than a sharp
truncation at the virial radius but keeps the outer part of the halo
in equilibrium.

We calculate the adiabatically compressed dark matter density profile
from the initial (NFW) dark matter profile and the final baryon
(Hernquist) profile under the assumptions of homologous contraction,
circular orbits for the particles, and conservation of angular
momentum $r\,M(<r)$:
\begin{equation}
  r_i\,[M_b(r_i)+M_{dm}(r_i)] = r_f\,[M_b(r_f)+M_{dm}(r_i)]\,. 
\label{adiabatic}
\end{equation}
Although it is not obvious that the adiabatic contraction
approximation should be directly applicable to elliptical galaxies --
it was originally proposed to describe the response of a dark matter
halo to baryonic dissipation during disk formation -- simulations have
shown it to be surprisingly accurate for modeling ellipticals.
Recent hydrodynamical simulations \citep{gkkn04} find that equation
(\ref{adiabatic}) overpredicts the amount of contraction at small
radii but this difference is much smaller than the difference between
haloes with and without adiabatic contraction, so we do not expect it
to significantly affect our results.

\begin{figure}
\includegraphics[scale=0.9]{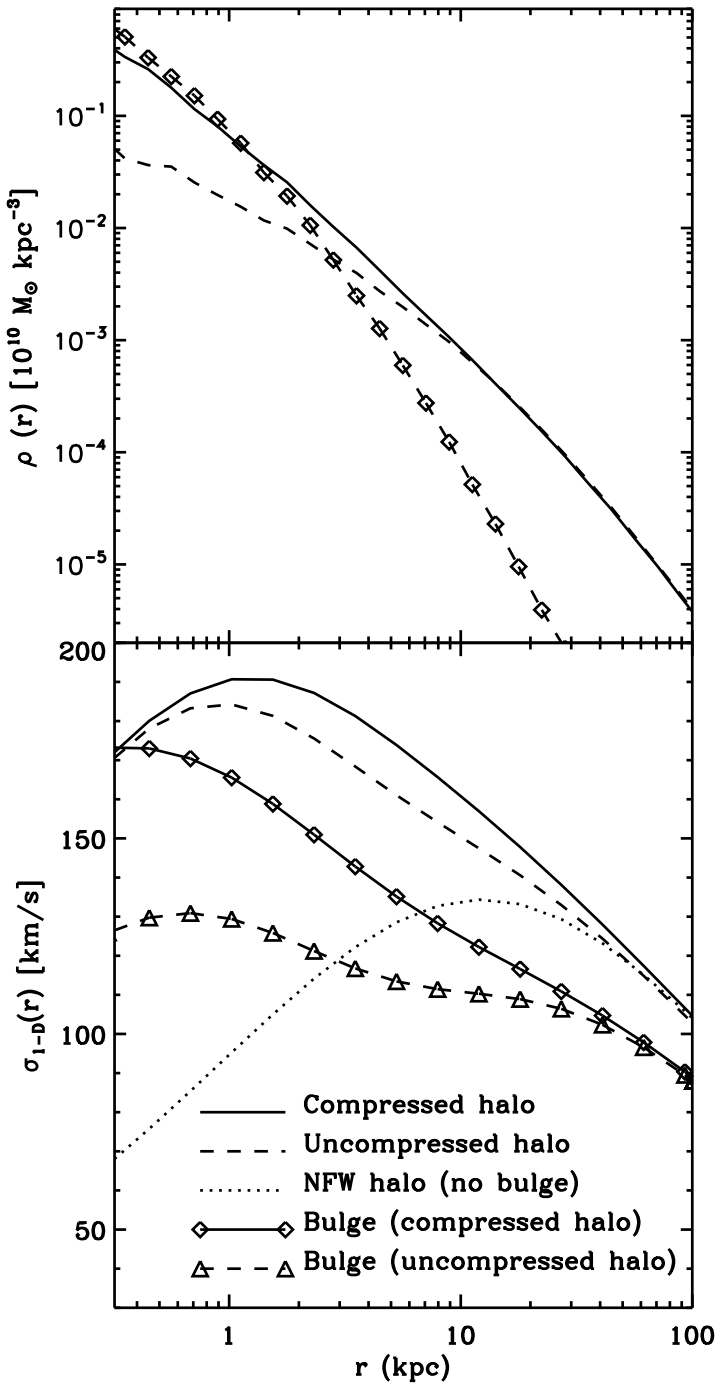}
\caption{
  Initial radial density profiles $\rho(r)$ (top) and 1-d velocity
  dispersion profiles $\sigma(r)$ (bottom) of the stellar bulge
  (symbols) and dark matter halo (plain curves) used in our
  simulations.  The dark matter and bulge masses are set to be
  $M_{dm}=10^{12} M_\odot$ and $M_*=M_{dm}/20$ here.  An adiabatically
  compressed dark matter halo (solid) has a larger $\rho$ and $\sigma$
  than the standard NFW halo (dashed) in the inner tens of kpc.  The
  velocity profile of a standard bulge-less NFW halo of the same mass
  is plotted (dotted curve) for comparison.  (The density profile of
  the bulge-less NFW halo is the same as the uncompressed NFW halo).}
\label{rhosig}
\end{figure}

Fig.~\ref{rhosig} shows the initial density profiles (top panel) and
1-d velocity dispersions (bottom panel) of the stellar bulge (symbols)
and dark matter haloes (plain curves) in our simulations.  It also
illustrates the effects of compression in the dark matter haloes (solid
vs dashed curves).  The top panel shows that the mass density interior
to $R_e= 2.8$ kpc is dominated by the stellar component when the dark
matter halo follows the NFW profile without compression, but
compression increases the central dark matter density significantly
and raises it to be larger than the stellar density at $r \ga 1$ kpc.
Compression also steepens the NFW $\rho\propto r^{-1}$ density cusp to
nearly isothermal ($r^{-2}$) at $r\ga 1$ kpc (although the compressed
density still asymptotes to $r^{-1}$ at very small $r$).  The bottom
panel shows that adiabatic compression increases the velocity
dispersion of both the dark matter and stellar components.  It is also
interesting to note that regardless of compression, the mere presence
of a stellar bulge in a dark matter halo increases the velocity
dispersion of the dark matter component by up to a factor of $\sim 2$
(for $r > 1$ kpc) in comparison with that of an NFW halo without a
bulge (dotted curve).  This effect also greatly reduces the radius at
which the ``temperature inversion'' characteristic of the NFW profile
occurs.

We construct the $N$-Body models by setting the bulge and dark matter
halo to be in equilibrium in the total gravitational potential.  We
assume that each component is both spherical and isotropic.  The
particle positions for each component are initialized from the density
profile of that component.  The particle velocities are drawn based on
each component's equilibrium distribution function $f_i(E)$, which we
calculate numerically using Eddington's formula \citep{bt87}:
\begin{equation}
  f_i(E)=\frac{1}{\sqrt{8} \pi^2} \int_0^E \frac{d^2 \rho_i}{d \psi^2}
  \frac{d \psi} {\sqrt{E-\psi}}\,,
\label{eqn-df}
\end{equation}
where $\rho_i$ is the density profile of component $i$ while $\psi$ is
the \emph{total} gravitational potential.  (This is to be contrasted
with the typical one-component model, where $\rho$ and $\psi$ are
related by Poisson's equation.)  This method of selecting the
velocities is self-consistent and makes no assumptions about the local
form of the velocity distribution, allowing us to create stable
initial conditions (see \citet{kmm04} for a full discussion of
sampling velocities when constructing initial conditions for N-body
models).  Note that while the \emph{positions} of the particles in
each component are the same as would be given in a one-component
model, the \emph{velocities} are not due to the presence of the other
component.

\subsection{Model Parameters and Merger Orbits}
\label{modelpara}

Table \ref{table-ICs} summarizes the parameters used in the galaxy
models and simulations for our series of production and test runs.
Before adiabatic compression, the initial haloes in our simulations
have $M_{dm}=10^{12} M_{\odot}$ and concentration $c=10$, giving a
scale radius of $r_s=16.26$ kpc and virial radius of $r_v=162.6$ kpc.
Gravity is the only physics in our simulation, meaning we can in
principle interpret the results at different mass scales by a
corresponding scaling of time and length.  In practice, however, the
presence of a stellar bulge complicates this scaling (see Section
\ref{section:discuss} for details).  We consider two values for the
total dark matter to stellar mass ratios, $M_{dm}/M_*=20$ ($M_*=5
\times 10^{10} M_{\odot}$) and $M_{dm}/M_*=10$ ($M_*=10^{11}
M_{\odot}$), which are chosen to be representative of the values
estimated from SDSS \citep{pad04} and weak lensing \citep{seljak02}.
As discussed in Section \ref{sec:properties}, we do not find the
merger remnants to depend sensitively on this ratio.  We relate the
remaining free parameter of the Hernquist density profile -- the
length scale $R_e$ -- to the stellar mass using the observed $\re-
M_*$ relation of early-type galaxies in SDSS
\citep{shen03}\footnote{Note the correct coefficient in Table 1 of
  \citet{shen03} is $2.88 \times 10^{-6}$ rather than $3.47 \times
  10^{-5}$ (S. Shen, private communication)}:
\begin{equation}
\label{eqn:shen}
    \re = 4.16 \left( \frac{M_*}{10^{11} \, M_{\odot}}
\right) ^{0.56} \, {\rm kpc}
\end{equation} 
Our choices of $M_{dm}/M_*$ and $\re$ do not guarantee that the
velocity dispersion $\sige$ of a model bulge will lie on the observed
\fpp.  In fact, observations of the \fpp\ measure the Faber-Jackson
$L-\sige$ relation while our simulations use the $M_*-\sige$ relation.
We relate the two using the $r^*$ band values from SDSS
\citep{bernardi2}.  Assuming $M_*/L=3-3.5$ in the $r^*$ band,
consistent with the estimates of Kauffmann et al. (2003), we find that
a bulge of $M_*=5 \times 10^{10}\, M_{\odot}$ should have a
characteristic velocity dispersion of $\sim 160$ km/s.
Table~\ref{table-ICs} shows that this is reasonably consistent with
the values we obtain from model M20 (which includes adiabatic
contraction) and is noticeably higher than the uncompressed halo
model (M20u).  In order to lie on the $M_*-\sige$ relation, the
uncompressed model must have a very large value of $M_*/L \sim 8$.
Adiabatic contraction (and the associated increase in the dark matter
fraction at small radii) therefore helps put the initial galaxy models
closer to the observed \fpp\ (see, however, \citealt{borriello03} for
an alternate view).

In addition to the internal structure of the bulge and dark matter
halo, the orbit of a binary encounter is also important in determining
the properties of the merger remnants.  For instance, it is well-known
that radial orbits tend to result in prolate remnants whereas orbits
with significant angular momentum generally give oblate or triaxial
end-products (\citealt{moore04} and references therein).  Most
merger studies to date have assumed parabolic orbits with somewhat ad
hoc impact parameters.  We attempt to use more realistic orbits by
drawing from recent results on the orbital distribution of merging
dark matter haloes in cosmological N-Body simulations.

\citet{benson04}, for example, has analyzed the orbital parameters of
minor mergers in the Virgo Consortium's N-Body simulations of a
$\Lambda$CDM model.\footnote{http://www.mpa-garching.mpg.de/Virgo/} He
finds that the orbital distribution peaks at semi-major axis $\approx
r_v$ and eccentricity $e$ slightly less than 1.  \citet{kb05} studied
similar quantities in major mergers (with mass ratio within 4:1) and
found that the distribution of circularities peaks at $\epsilon
=\sqrt{1-e^2}=0.5$, consistent with the findings of Benson, an earlier
study by \citet{tormen97}, and recent results from \citet{zentner05}.
We therefore choose a ``most probable'' orbit defined by
$\epsilon=0.5$ and pericentric distance $r_p=50$ kpc, consistent with
Khochfar \& Burkert's relation $r_p = 210 \epsilon^{2.07}$ kpc.  We
assign the radial and tangential velocities $v_r$ and $v_{\theta}$ of
this orbit to agree with Benson's analysis (his Figs.~2 \& 7, and
equation~2): $v_r=v_{vir}$ and $v_{\theta}=0.7 v_{vir}$ where
$v_{vir}^2=GM_{vir}/R_{vir}$.  In addition to the ``most probable''
orbit, we also simulate a head-on parabolic merger for comparison.
Finally, we perform one modified ``most probable'' simulation in which
both $v_r$ and $v_{\theta}$ are reduced by 15\% in order to test how
robust our results are to the choice of initial orbital velocities.
The initial center-of-mass velocity for each run is listed in
Table~\ref{table-ICs}.  In all cases we start with the galaxy centers
separated by two virial radii.

\subsection{Simulations and Resolution Studies}
\label{sec:sim_and_res}
We use {\tt GADGET} \citep{gadget}, an efficient parallelized tree
code, to perform the simulations presented below.  We set the force
softening $\epsilon=0.3$ kpc for both star and dark matter particles;
this choice corresponds to either $0.072 \re$ ($M_*=10^{11}
M_{\odot}$) or $0.11\re$ ($M_*=5 \times 10^{10} M_{\odot}$) for our
initial galaxy models.  We run each model until the remnant reached
virial equilibrium (between 4 and 6 Gyr depending on the orbit).
Energy is conserved to within 0.5\% for all runs and the global virial
ratio $2T/|W|$ is unity to within 1.5\% at the final timestep in each
case.

\begin{table}
\caption{Simulation and model parameters of each
galaxy in the initial conditions used in our production runs (upper 5)
and test runs.  In the run names, the merger orbits are distinguished
by ``M'' and ``R'' for ``most probable'' vs. radial (parabolic)
orbits; the mass fraction $M_{dm}/M_*$ is labeled by 20 vs. 10; the
runs with initially uncompressed dark matter halos are denoted by
``u''.  Run N20 is identical to run M20 except the magnitude of each
component of the initial orbital velocity is reduced by 15\%. 
$N_{dm}$ and
$N_*$ are the numbers of dark matter and stellar particles used in
each simulation.  All dark matter halos have virial masses of
$10^{12}~M_{\odot}$ and virial radii of 162.6 kpc.} 
\label{table-ICs}
\begin{tabular}{lllllll}
\hline
\hline
\textbf{Run} &$\mathbf{\frac{M_{dm}}{M_*}}$ &$\mathbf{N_{dm}}$
&$\mathbf{N_*}$ &$\mathbf{\re}^a$ &$\mathbf{\sige}^b$ &$v_{CM}^c$\\
\hline
M10  &10 &$5.5 \times 10^5$ &49000 &4.13 &173.5 &200\\
M20   &20 &$5.5 \times 10^5$ &24500 &2.85 &151.4 &200\\
N20  &20 &$5.5 \times 10^5$ &24500 &2.85 &151.4 &170\\
M20u &20 &$5.5 \times 10^5$ &24500 &2.83 &122.7 &200\\
R20   &20 &$5.5 \times 10^5$ &24500 &2.83 &150.4 &250\\
\hline
M10t1    &10 &$1.1 \times 10^5$ &49000 &4.13 &172.8 &200\\
M20t0    &20 &$1.1 \times 10^5$ &4900 &2.79 &151.3 &200\\
M20t1    &20 &$1.1 \times 10^5$ &24500 &2.76 &150.9 &200\\
M20t2    &20 &$1.1 \times 10^5$ &49000 &2.80 &150.1 &200\\
M20t1u   &20 &$1.1 \times 10^5$ &24500 &2.76 &123.1 &200\\
R20t1    &20 &$1.1 \times 10^5$ &24500 &2.76 &150.9 &250\\
\hline
\end{tabular}

\medskip
$^a$effective radius of stellar bulge, in kpc\\
$^b$aperture velocity dispersion of stellar bulge (eq.~\ref{sige}), 
in km s$^{-1}$\\
$^c$center-of-mass velocity for merger, in km s$^{-1}$\\
\end{table}

We have performed several convergence tests to ensure that sufficient
particle numbers and force resolution are used for both the dark
matter and stellar components.  We find that increasing the number of
star particles $N_*$ from 24500 to 49000 (runs M20t1 and M20t2) at
fixed dark matter particle number does not affect our results for
$\re$ and $\sige$.  It is, however, important to use $N_{\rm
  dm}\approx 5 \times 10^5$ dark matter particles per halo since our
test runs with $N_{\rm dm}\approx 10^5$ show noticeable differences:
for example, if we assume $\re \propto M_*^{\alpha}$, using $N_{\rm
  dm}=10^5$ rather than $5 \times 10^5$ causes an overestimation of
$\alpha$ at the 15\% level (for a compressed halo with $M_{dm}/M_*=20$).
This effect can be understood in terms of two-body relaxation: using a
small number of particles means that on small scales, individual
particle interactions become important even when compared to the mean
gravitational field.  This causes an artificial heating, diluting the
dark matter potential at the center of the model galaxy, and resulting
in an artificially expanded stellar bulge.  Similarly the final
stellar velocity dispersions in the test runs tend to come out lower
than in the production runs, again because of the shallower central
potential.  This effect leads to an over-estimate of $\re$ and an
under-estimate of $\sige$, biasing the \fpp\ correlations we hope to
measure.  As listed in Table~\ref{table-ICs}, our production runs all
use $5.5\times 10^5$ dark matter particles and 24500 stellar particles
(for $M_{dm}/M_*=20$).  These values provide enough mass resolution in
both dark and stellar matter to resolve quantities of interest on
scales larger than $\sim 1$ kpc.

\section{Properties of Merger Remnants}
\label{sec:properties}
The global merger dynamics in our simulations using the ``most
probable'' orbit is similar in all three cases (runs M10, M20, and
M20u).  As the galaxies spiral in toward each other, the dense stellar
bulges stay intact while the more diffuse dark matter haloes mix
rapidly.  Dynamical friction then brings the bulges and the central
core of the dark matter haloes together and the merger is finished
after approximately 5 Gyr.  
The process for the radial orbit is about 30\% faster even though the
initial orbit is more unbound because the orbit brings the bulges
together earlier.
In the following sections we quantify the structure and kinematics of
the merger remnants and then discuss the extent to which the merger
preserves homology and maintains the fundamental plane relations of
elliptical galaxies.

\begin{table*}
\caption{Fits to the radial density profiles of dark
matter haloes and stellar bulges at $t=0$ and 6 Gyr in our five
production runs.  Parameters $A$, $B$, and $C$ are defined in
equation~(\ref{eqn:rho_fit}).  Haloes are fit on the range [1,100] kpc,
bulges on the range [0.32,16] kpc.}
\begin{tabular}{lccccccccccc}
\hline
\hline
& &\vrule &\multicolumn{4}{|c|} {\bf Dark Matter Haloes} &\vrule
&\multicolumn{4}{|c|}{\bf Stellar Bulges}\\  

\textbf{Run} &time &\vrule & $M_v$ & $A$ & $B$ & $C$ &\vrule & $M_*$
&$r_{1/4}$ & $B$ & $C$ \\

   & (Gyr) & \vrule & ($10^{10} M_\odot\!$)  & ($10^5 M_{\odot} 
   \mathrm{kpc}^{-3}\!$) & (kpc) &  &\vrule  
  &$(10^{10}M_\odot$)   & (kpc) & (kpc) &\\

\hline
M10  &0 &\vrule &100.6 &14.8 &32.5 &1.8  &\vrule &10.1 &2.24 &2.19 &0.97\\ 
        &6 &\vrule &133.8 &17.3 &34.1 &1.8  &\vrule &18.6 &2.94 &4.47 &1.5\\ 
M20   &0 &\vrule &100.4 &9.52 &39.4 &1.8  &\vrule &5.02 &1.57 &1.55 &0.98\\ 
        &6 &\vrule &136.8 &12.1 &40.2 &1.8  &\vrule &9.75 &2.26 &3.83 &1.6\\
N20 &0 &\vrule &100.4 &9.52 &39.4 &1.8  &\vrule &5.02 &1.57 &1.55&0.98\\ 
    &5.5 &\vrule &144.1 &9.87 &44.8 &1.8  &\vrule &9.82 &2.21 &4.31 &1.7\\
M20u &0 &\vrule &100.0 &134  &15.8 &0.98 &\vrule &5.08 &1.56 &1.68 &1.1\\ 
        &6 &\vrule &136.2 &114  &18.8 &1.1  &\vrule &9.92 &2.04 &3.88 &1.7\\ 
R20   &0 &\vrule &100.4 &6.84 &45.2 &1.8  &\vrule &5.06 &1.56 &1.55 &0.99\\ 
        &4 &\vrule &132.1 &14.6 &37.6 &1.7  &\vrule &10.1 &2.89 &7.31 &1.9\\ 
\hline
\end{tabular}
\label{table-rho} 
\end{table*}

\subsection{Surface Brightness and Radial Density Profiles} 

\begin{figure}
\includegraphics[scale=0.53]{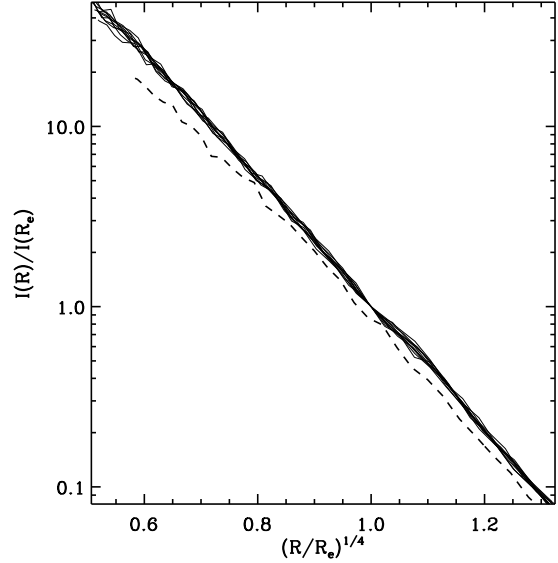}
\caption{Surface brightness profiles for run M20 at t=0 Gyr (dashed
curve) and ten random projections at t=6 Gyr (solid curves) as a
function of $(R/\re)^{1/4}$ and normalized to unity at $R=\re$ (the
initial profile is offset by 15\% for clarity).  A de Vaucouleurs law
is a straight line on this plot.}
\label{figure:surfb}
\end{figure}

One distinctive property of the bulges in early-type galaxies is that
their surface brightness profiles follow the de Vaucouleurs law, $\log
I(R) \propto -R^{1/4}$.  Fig.~\ref{figure:surfb} shows $I(R)$ for 10
random projections of the final bulge from run M20 (solid lines) as a
function of $(R/\re)^{1/4}$.  For comparison we also show $I(R)$ of
the initial bulge (dashed line; offset by 15\% for clarity).  Each of
the random projections of the final bulge, as well as the initial
profile, all follow the de Vaucouleurs law over the plotted range
of $\epsilon \le R \le R_{90}$ (where $R_{90}$ is the projected radius
containing 90\% of the bulge mass). 

Although the remnants' projected density profiles mimic those of the
initial conditions, the unprojected density profiles are not
necessarily identical.  In order to quantify the changes in density
structure induced by the merger, we fit both the bulges and dark
matter haloes to the following profile:
\begin{equation}
   \rho(r)= {A\over (r/B)^C (1+r/B)^{(D-C)}}.
\label{eqn:rho_fit}
\end{equation}
Since we find the outer slopes of both the dark matter haloes and
stellar bulges to be mostly unaffected by the merger, we fix $D$ to
their initial values: $D=3$ for the halo and $D=4$ for the stellar
bulge.  Models with $D=4$ have been investigated by \cite{dehnen93}
and \cite{tremaine94}; for these it is possible to relate $A$ to
$B,~C, \mbox{ and } M_*$: $A=(3-C)M_*/(4 \pi B^3)$.  Initially the
bulges are all described by Hernquist density profiles, so $A=M_*/(2
\pi a^3)$, $B=a=r_{1/4}$, $C=1$, and $D=4$.  The adiabatically
compressed dark matter haloes differ from the original NFW profiles, so
we also fit the initial compressed halo profiles to
equation~(\ref{eqn:rho_fit}) with $D=3$.

The results are summarized in Table~\ref{table-rho} and show an
interesting trend: the density profiles of the dark matter haloes
retain their shape to a remarkable degree throughout the merger.  In
fact, the main difference between the initial and final $\rho_{dm}$ is
an overall amplitude change by a factor of $\sim 1.4$, regardless of
the merger orbit or whether the halo is compressed.  The bulges show
more evolution after the mergers, especially in the inner slope of the
density profile.  In all cases the fitted inner slope of the bulge
steepens noticeably.  Although observations indicate that massive
elliptical galaxies have surface brightness cores \citep{faber97}, the
size of these cores tends to be $\sim 100$ pc (for the galaxy masses
in our simulations).  
This scale is below our force softening of
$\epsilon=0.3$ kpc, so there is no inconsistency between our surface
brightness or density profiles and observations of elliptical
galaxies.  As with the dark matter, the total profile for stars plus
dark matter changes mostly by an overall amplitude and remains close
to isothermal on the scale of $R_e$.

We also calculate the final virial radius of each dark matter halo and
compare the enclosed mass $M_v$ with the initial value in
Table~\ref{table-rho}.  In each case only about 65-70\% of the initial
virial mass ends up in the final virial radius; a similar ratio was
found in our earlier study of dark-matter-only mergers \citep{mrbk04}.
This ``puffing up'' of the outer part of the dark matter halo is in
stark contrast to the stellar bulge, where 
essentially all of the initial stellar mass $M_*$ ends up in the final
bulge.  (This result is insensitive to changes in merger orbit,
compression of halo, and the value of $M_{dm}/M_*$.)  The ratio of the
dark matter to stellar mass within the virial radius therefore
decreases after a major collisionless merger.  However, as discussed
below in Section \ref{section:dmfraction}, the dark matter-to-stellar
mass ratio within the central region of a galaxy (within $\re$)
shows the opposite behavior.

\subsection{Remnant Shapes}
\label{shape}

\begin{figure}
\begin{center}
\includegraphics[scale=0.45]{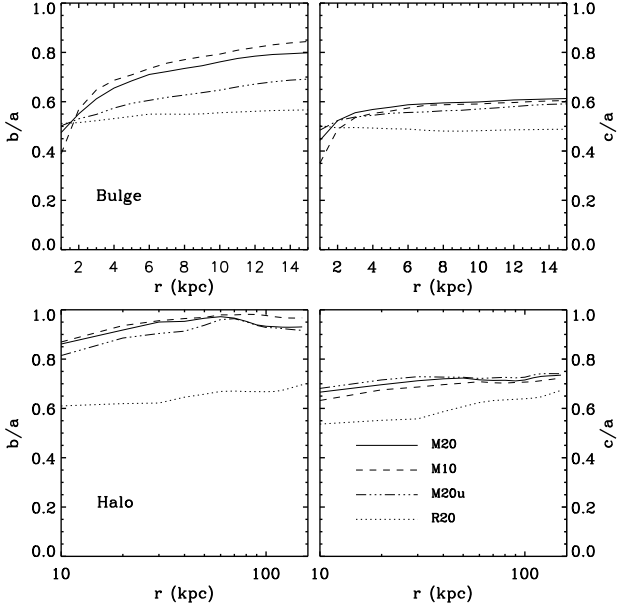}
\end{center}
\caption{Triaxiality of the merged stellar bulges (top) and dark
matter haloes (bottom) as a function of radius at $t=6$ Gyr from 
four production runs.  Axis ratios $b/a$ (left) and $c/a$ (right) are
shown ($a\ge b \ge c$ by definition).  Radial orbits (dotted curves)
tend to produce prolate bulges and haloes, whereas orbits with non-zero
angular momentum result in triaxial bulges and nearly oblate haloes.
Note that the ratios for the bulge are plotted linearly in radius,
while those for the halo are plotted logarithmically.}
\label{figure-axis}
\end{figure}

We find that the merger remnants in our simulations tend to be
triaxial, with their structure determined by the initial orbit: radial
orbits produce nearly prolate remnants while orbits with non-zero
angular momentum yield triaxial products.

To quantify this trend, we determine the axis ratios of the final
products using an iterative eigenvalue technique \citep{dc91}.  We
diagonalize a modified inertia tensor $I_{ij}=\sum x_i x_j/R^2$ (where
$R^2=x^2+y^2/(b/a)^2+z^2/(c/a)^2$ is the ellipsoidal radius of a
particle and $a \ge b \ge c$ are the eigenvalues of $I_{ij}$) by first
assuming spherical symmetry and computing $I_{ij}$, then using the
eigenvalues of iteration $n$ as input values for iteration $n+1$.  We
also compare the axis ratios calculated both cumulatively ($r_i \le
R$) and in differential shells ($R_1 \le r_i \le R_2$).  For the dark
matter haloes, these two method give similar results at most radii, but
near $r_v$ the cumulative method is less sensitive to local changes in
the axis ratios due to the contribution of many particles at smaller
radii, an effect noted previously in \citet{kazan04}.  The cumulative
method is, however, more robust for the bulges due to the relatively
low number of stellar particles.  For the following analysis we thus
use the cumulative method for measuring the shapes of both components.

The top two panels of Figure~\ref{figure-axis} show the axis ratios of
the remnant stellar bulges at $t=6$ Gyr from 
four of our production runs (we do not include run N20 because its
axis ratios are nearly indistinguishable from run M20).  The radial
orbit (dotted curves) results in a prolate bulge with both $b/a$ and
$c/a \approx 0.5$, irrespective of radius.  The ``most probable''
orbit, on the other hand, yields a triaxial bulge with axis ratios
that have a mild radial dependence.  These axis ratios are in
agreement with a recent analysis of the SDSS data that found that
bulges of early-type galaxies are consistent with triaxial ellipsoids
\citep{vr05}, though we note that \citet{alam02} also found that
prolate ellipsoids are consistent with the SDSS early data release
sample.  Simulations of a brightest cluster galaxy from a cosmological
simulation \citep{dubinski98} find a stellar remnant with $b/a =0.66$,
which is similar to our remnants of the ``most probable'' orbits, and
$c/a=0.47$, which is closer to what we find for a radial merger.  The
latter is a consequence of the correlated directions of Dubinski's
multiple mergers along filaments.
 
The bottom two panels of Figure~\ref{figure-axis} show the axis ratios
for the merged dark matter haloes.  The radial orbit again results in a
mostly prolate halo, whereas the three ``most probable'' orbit runs
give nearly oblate haloes with $b\approx a$ and $c/a \approx 0.7$,
largely independent of $M_{dm}/M_*$ and halo compression.  It is
useful to compare our results with those from dark-matter-only
cosmological simulations of \citet{bs04}.  For a fair comparison, we
calculate the axis ratios of our remnants in a shell from 0.25 to
$0.4\,r_v$, the same range as Bailin and Steinmetz use.  We find the
``most probable'' orbit to give $b/a \approx 0.9-0.95$ and $c/a
\approx 0.7$ while the radial orbit gives a remnant with $b/a \approx
c/a \approx 0.7$.  Both agree with the results of \citet{js02} and
Bailin \& Steinmetz (cf their fig. 5), though our values of $b/a$ tend
to be at the high end of their distribution.

The difference in the ``most probable'' orbit runs between the bulge
(triaxial) and halo (oblate) shapes is noteworthy, as it relates to
the difference between the merging of the haloes and of the bulges.
The haloes merge relatively quickly, but the bulges circle each other
and spiral inward due to dynamical friction on the haloes.  The final
plunge of the bulge tends to be relatively radial even though the
initial orbit has substantial angular momentum, giving the remnant
bulge a different shape from the remnant haloes.  Quantitatively, we
find that the amount of (initial orbital) angular momentum lost from
the bulge is $\sim 75$\% (i.e.  $\vec{L}_f \approx 0.25 \vec{L}_i$).
As a related measure, we calculated the ratio of rotation velocity to
velocity dispersion for several projections.  As might be expected,
the maximal $v_{rot}/\sigma \approx 0.5$ at $\re$ occurs in the plane
of the orbit, while for most lines of sight $v_{rot}/\sigma < 0.3$,
typical of a slowly rotating, pressure-supported bulge.  This is
consistent with recent simulations by \citet{nb03}, who found that
equal mass mergers of disk galaxies tend to result in
pressure-supported ellipticals while mergers with mass ratios of 3:1
and 4:1 result in oblate rotators (with $v_{rot}/\sigma > 0.7$).

\subsection{The Fundamental Plane Relations}

Before investigating the scaling relations among the effective radius
$R_e$, velocity dispersion $\sigma_e$, and mass of the stellar bulge
$M_*$, we first describe how we determine $R_e$ and $\sigma_e$.
Because the remnant bulges are triaxial, their observable properties
depend on the angle from which the bulge is viewed.  To quantify this
effect, we calculate the properties of each of our remnant bulges for
$10^4$ random viewing angles.  A standard technique for obtaining the
effective radius is to fit the projected profile to a Sersic profile
and use the derived $\re$ as the half-light radius.
As described in Appendix \ref{section:re}, however, we find this
procedure does not give unique results, so we employ an alternate
technique here.  We calculate the surface mass density perpendicular
to each line of sight in spherical bins as well as the projected
enclosed mass at each radius.  We define the projected radius of the
particle enclosing half the total mass to be the half-light radius
$\re$.  This procedure recovers the correct $\re$ for the initial
Hernquist profile to within 1-2~\%.  For the velocity dispersion, we
use the surface-brightness weighted dispersion within $\re$,
\begin{equation} 
     \sige^2 \equiv \sigma_a^2(\re) = \frac{\int_0^{\re}
     \sigma^2_{los}(R) I(R) R dR} {\int_0^{\re} I(R) R dR},
\label{sige} 
\end{equation} 
where $I(R)$ is the surface brightness and $\sigma_{los}$ is the
line-of-sight velocity dispersion.  Observationally, a commonly used
definition is the dispersion measured within an aperture of size
$\re/8$.  For our simulations, however, $\re/8$ is comparable to the
force softening $\epsilon$, so we instead define our velocity
dispersion within $\re$.  We do not expect this modification to bias
our results with respect to observations because the aperture
dispersion profile $\sigma_a(R)$ is generally quite flat within $\re$
\citep{gerhard01}.  Any error introduced will almost certainly be less
than that associated with empirically correcting the observed aperture
dispersions to $\re/8$.  To be conservative with respect to force
accuracy, we use $2 \epsilon$ as the minimum radius for the integral
in equation (\ref{sige}).

\begin{figure}
\begin{center}
\leavevmode
\includegraphics[scale=0.55]{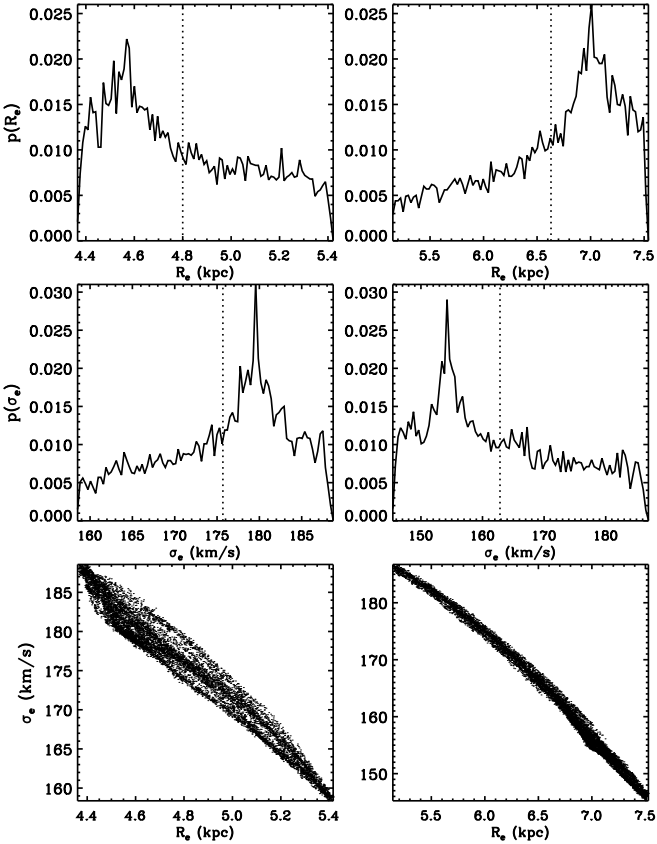}
\end{center}
\caption{Distributions of $\re$ (top) and $\sige$ (middle), as well as
  the correlation at fixed mass between $\re$ and $\sige$ (bottom),
  for 10000 random viewings of the merged stellar bulge in runs with
  $M_{dm}=20 M_*$.  Left: ``most probable'' orbit with $v_{\theta}/
  v_r \approx 0.7$ (run M20).  Right: radial orbit (run R20).  Note
  that the mean (dotted line) and mode can differ significantly, as
  the distributions are heavily skewed.  The distributions are also
  dependent on orbit: for run M20 the mean $\re$ is greater than the
  mode and the mean $\sige$ is smaller than the mode, while the
  reverse holds for run R20.  In both cases the ``observed'' $R_e$ and
  $\sige$ are highly correlated, as indicated by the bottom two
  panels.}
\label{histmulti}
\end{figure}

\begin{table*}
\caption{Mean (left) and mode (right) of the distributions of the
effective radius $R_e$ and velocity dispersion $\sige$ from 10000 random
projections of the remnant bulge in each of our five production runs.
Parameters $\alpha$ and $\beta$ are the resulting exponents for two
projections of the fundamental plane relation: $\re \propto
M_*^{\alpha}$ and $M_* \propto \sige^{\beta}$.} 
\begin{tabular}{llcclllccll}
\hline
\hline
&\vrule & &Mean & & &\vrule & &Mode & &\\
\textbf{Run} &\vrule &$\mathbf{\re}$ &$\mathbf{\sige}$
&$\mathbf{\alpha}$ &$\mathbf{\beta}$ &\vrule &$\mathbf{\re}$
&$\mathbf{\sige}$  &$\mathbf{\alpha}$ &$\mathbf{\beta}$\\ 
  & \vrule & (kpc) & (km/s) &  &  & \vrule & (kpc) & (km/s) & & \\
\hline
M10  &\vrule &6.94 &200.0 &0.75 &4.9 &\vrule &6.45 &204.5 &0.64 &4.2\\  
M20  &\vrule &4.80 &175.7 &0.75 &4.7 &\vrule &4.58 &179.0 &0.69 &4.1\\   
N20  &\vrule &4.85 &173.0 &0.77 &5.2 &\vrule &4.63 &176.0 &0.70 &4.6\\
M20u &\vrule &4.53 &153.0 &0.68 &3.1 &\vrule &4.49 &150.0 &0.67 &3.5\\ 
R20  &\vrule &6.63 &162.8 &1.23 &8.7 &\vrule &7.01 &154.3 &1.31 &27\\    
\hline
\end{tabular}
\label{table-runs} 
\end{table*}

Figure \ref{histmulti} shows the resulting distributions of $\re$ (top
panels) and $\sige$ (middle panels) from $10^4$ random viewing angles
of our merger remnants from two simulations with different orbits:
``most probable''(left panels) and radial (right panels).  The
distributions are highly non-Gaussian, although in each case there is
a reasonably well-defined mode that differs significantly from the
mean (indicated by the vertical dotted line).  The two orbits result
in strikingly different distributions of $\re$ and $\sige$.  This
difference is primarily because the two classes of orbits yield
remnants with different shapes (see Section \ref{shape} above).  The
``most probable'' orbit yields a triaxial remnant for which one is
more likely to observe a small $\re$ whereas the radial orbit yields a
prolate remnant for which one is more likely to observe a large $\re$.
In addition, mergers on orbits with non-zero angular momentum (such as
the ``most probable'' orbit) suffer more dynamical friction and thus
more energy transfer from the stellar bulge to the dark matter (see
also Section \ref{homology} and Table~\ref{table-homology}).  This
effect leads to a smaller and more tightly bound bulge with a larger
velocity dispersion.

The bottom panels of Figure~\ref{histmulti} show the correlation
between the \fpp\ parameters $\re$ and $\sige$ for fixed stellar mass
from the $10^4$ viewing angles.  It is evident that the measurements
are highly correlated; viewing angles that yield a relatively small
effective radius result in a relatively large velocity dispersion.
This effect is due to projection: looking down the major axis of a
prolate remnant, for example, leads to a smaller effective radius and
larger velocity dispersion than if the viewing angle is along a minor
axis.  These correlations are also expected from the fundamental plane
at fixed luminosity (or mass): from the SDSS values for
equation~(\ref{eqn:fp}), $\re \propto \sige^{-3}$ at fixed luminosity.
Although the fundamental plane cuts through the $\re-\sige$ plane in a
non-trivial manner, making a direct comparison between the spread in
our $\re-\sige$ correlations and the scatter of the fundamental plane
difficult, the thickness of the correlations we present here are
consistent with the observed small scatter in the fundamental plane.
However, it is not clear how additional effects, such as scatter in
the global $M_{dm}/M_*$ and in initial sizes and shapes of merging
galaxies, would affect this scatter.

Using the results of Figure~\ref{histmulti}, we can now assess the
extent to which the stellar bulges remain on the projections of the
fundamental plane after major dissipationless mergers.
Table~\ref{table-runs} lists the mean and mode of $\re$ and $\sige$
for the merged stellar bulges and the slopes for power-law fits to
$\re-M_*$ and $M_*-\sige$ from our five production runs.  Since the
mean and modes differ significantly, we present results for each
definition of the ``average'' remnant.  The results for $\re-M_*$
shown in Table~\ref{table-runs} are more robust than those for
$M_*-\sige$ because of the steepness of the latter relation.

Table~\ref{table-runs} shows that for the ``most probable'' orbit with
adiabatically compressed dark matter haloes (i.e. runs ``M10'' and
``M20'', which are the best motivated models studied in this paper),
the derived \fpp\ slopes using the mode are reasonably consistent with
observations:
\begin{eqnarray}
\label{eqn:fpslopes}
  && \re \propto  M_*^{\alpha}\,, \quad \alpha=0.64-0.69 \,, \nonumber\\
  && M_* \propto  \sige^\beta\,, \quad  \beta=4.1-4.2  \,.
\end{eqnarray}
Recall that the initial values of $\re$, $M_*$, and $\sige$ for these
simulations also lie on the \fpp\ (see Section \ref{modelpara}).  In
comparison, the ``most probable orbit'' model with an {\it
  uncompressed} dark matter halo (run ``M20u'') produces a $M_*-\sige$
relation somewhat shallower than the observed relation, while radial
orbits (run ``R20'') yield a remnant well off the \fpp \ ($\beta \sim
27$!).  In fact, the remnant from the head-on merger is much further
off the \fpp\ than the intrinsic scatter of the plane, suggesting that
few major mergers are radial or nearly so.  This inference is
consistent with the low frequency of head-on mergers in the orbital
analyses discussed in Section \ref{modelpara}.

\subsection{Homology and Energy Conservation}
\label{homology}

Semi-analytic galaxy formation models often use energy conservation
arguments and rely on the assumption of homology for predicting how
elliptical galaxies evolve during dissipationless mergers
\citep{cole00,shen03}.  The energy conservation equation is usually
derived using the fact that on dimensional grounds we can quantify the
energy of a single bulge of mass $M_*$ as
\begin{equation}
\label{eqn:bulgeE}
    E \equiv -f \frac{GM_*^2}{R_*} 
\end{equation} 
for some characteristic radius $R_*$ and structural parameter $f$ that
depends on the properties of the galaxy model under consideration.
For a bulge with the Hernquist profile (without a dark matter halo),
$E=-GM_*^2/12a$, so $f=\re/12a=0.151$ using $R_*=\re$, or
$f=(1+\sqrt{2})/12=0.201$ using $R_*=r_{1/2}$.  For our various
bulge-plus-halo models, we find $f=0.265-0.367$ (see
Table~\ref{table-homology}) due to the presence of a dark matter halo.

\begin{table}
\caption{Parameters from the bulge energy
conservation equation (eqn.~\ref{energy}) derived from the
simulations.  We quote numbers for the remnant bulge structural
parameter $f_f$ using both the mean $\re$ and mode $\re$ obtained from
$10^4$ random projections (listed in Table~\ref{table-runs}).
Parameters $f_{orb}$ and $f_t$ characterize the initial
orbital energy (at virial radius) and the energy
transfer between the stellar and dark matter components due to
mergers.}
\begin{tabular}{llllll}
\hline
\hline
\textbf{Run} &$f_i$ &$f_{f,mean}$ &$f_{f,mode}$ &$f_{orb}$ &$f_t$\\
\hline
M10  &0.342 &0.385 &0.358 &0.118 &0.344\\
M20  &0.367 &0.444 &0.424 &0.144 &0.469\\
N20  &0.367 &0.448 &0.428 &0.215 &0.394\\
M20u &0.265 &0.319 &0.316 &0.144 &0.375\\
R20  &0.367 &0.489 &0.517 &0.00715 &0.183\\ 
\hline
\end{tabular}
\label{table-homology} 
\end{table}

Using equation~(\ref{eqn:bulgeE}), we can write an energy conservation
equation for the bulges in a binary merger of two galaxies (with
initial bulge masses $M_1$ and $M_2$) as
\begin{equation} 
    f_f \frac{M_f^2}{R_f} = f_1 \frac{M_1^2}{R_1}+ f_2
    \frac{M_2^2}{R_2} + (f_{orb}+f_t) \frac{M_1 M_2}{R_1+R_2} \,,
\label{energy} 
\end{equation}
where $f_{orb}$ contains information about the orbital energy and
$f_t$ describes the energy transfer between the stellar and dark
matter components.  With this definition, $f_t>0$ implies that the
final bulge is more tightly bound than the initial bulges, i.e. energy
is transferred from the bulge to the dark matter halo.  (An equation
analogous to equation~(\ref{energy}) can be used to describe the dark
matter component.)  As noted by previous authors, setting
$f_f=f_1=f_2$ and $f_{orb}=f_t=0$ (i.e., perfect homology and a
parabolic orbit with no energy transfer between the dark matter and
the stellar component) implies $R \propto M_*$ for an equal mass
merger, which does not match the observed relation.  Alternatively,
assuming homology is satisfied ($f_f=f_i$) and $R \propto M_*^\alpha$,
one can derive that
\begin{equation}
  \frac{f_{orb}+f_t}{f_i} =\left(\frac{\xi^{\alpha}+1} {\xi} \right)
  \left[(\xi+1)^{2-\alpha} -\xi^{2-\alpha}-1 \right] \,.
\label{eqn:homology}
\end{equation}
for binary mergers between galaxies with mass ratio $\xi$.
The observed $\re-M_*$ relation is $\alpha=0.56$, implying
$f_{orb}+f_t=1.43 f_i$ for equal-mass mergers if homology is
satisfied.

To assess homology in our simulations we calculate the values of $f$
using the energy defined in Appendix \ref{appendix:energy}:
\begin{equation}
\label{eqn:Econs}
    E=\sum_i \left(\frac{1}{2}m_i v_i^2-\sum_{j > i} \frac{Gm_i
    m_j}{r_{ij}} -\sum_k Gm_i m_k \frac{\vec{r}_i \cdot
    \vec{r}_{ik}}{r_{ik}^3} \right)
\end{equation}
where $r_{ij} \equiv |\vec{r}_i-\vec{r}_j|$ and $i$ and $j$ run over
all stellar particles while $k$ runs over all dark matter particles.
The resulting values for $f$ are shown in Table~\ref{table-homology}.
Homology is preserved at the $\sim 15$\% level in all three runs with
the ``most probable orbit'' (runs M10, M20, and M20u): $f$ increases
slightly in these runs, with the mean value changing more than the
mode, but in each run both values are within 20\% of the initial
value.  The same is true for run N20, which is identical to M20 aside
from having lower orbital velocity.  The radial merger (run R20),
however, breaks homology significantly more than the ``most probable''
orbit.  In all five production runs, the ratio of bulge-dark matter to
bulge-bulge potential energy increases, reflecting the relative
distribution of dark matter and stars, a point which we discuss
further in Section \ref{section:dmfraction}.

Table~\ref{table-homology} also gives the values of $f_{orb}$ and
$f_t$, the parameters characterizing the amount of initial orbital
energy (at the virial radius) and the amount of energy transferred
between the stellar and dark components.  For the ``most probable''
orbit, $f_t \approx 0.34-0.47$, indicating that there is a significant
amount of energy lost from the stellar bulge to the dark matter halo
during the merger (partially via dynamical friction).  The energy
transfer is significantly less for the head-on merger in which
dynamical friction is less important.  The values of $(f_t +
f_{orb})/f_i \approx 1.3-1.7$ for the most probable orbit are
reasonably consistent with the estimate above for the amount of energy
transfer required to explain the observed $\re-M_*$ relation.
Interestingly, a comparison of run N20 to M20 show that while N20 has
less orbital energy, it also less energy transferred from the bulge to
halo and the net result is a halo with nearly identical homology
properties to those from run M20. 

\subsection{The Dark Matter Fraction}
\label{section:dmfraction}

\begin{figure}
\includegraphics[scale=0.53]{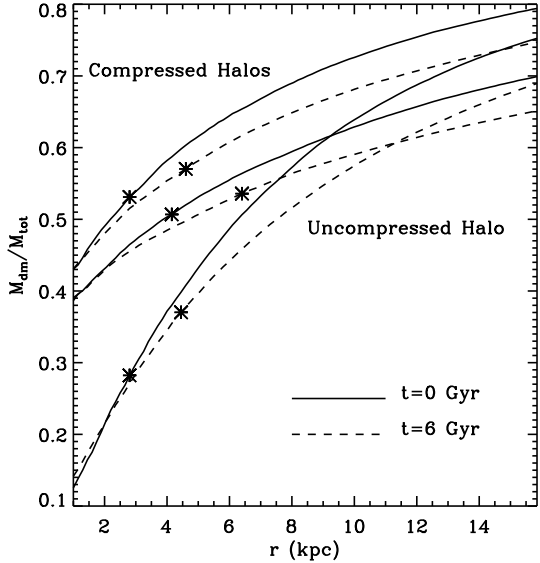}
\caption{Ratio of dark matter mass to total mass interior to radius
  $r$ as a function of $r$ for initial condition (solid curve) and
  remnant system (dashed curve) for simulation M20 (compressed halo;
  upper curves), M10 (compressed halo; middle curves), and M20u
  (uncompressed halo; lower curves).  The symbols mark the effective
  radius $R_e$ for the given simulation.  In each case the fraction of
  dark matter interior to $\re$ increases after the merger.}
\label{btodmratio}
\end{figure}

As noted in Section \ref{sec:intro}, the fundamental plane relations are
tilted from the simple one-component predictions based on the virial
theorem.  One possible explanation for the origin of this tilt is a
luminosity-dependent dark matter fraction within the effective radius.
In this section we test this hypothesis by quantifying how the dark
matter fraction evolves during our simulations.

In Figure~\ref{btodmratio} we plot the ratio of dark matter mass to
total mass $M_{dm}(<r)/M_{tot}(<r)$ as a function of radius for the
three simulations using the ``most probable'' orbit: runs M20 (upper
curves) and M10 (middle curves) have compressed dark matter halos
while run M20u (lower curves) has a standard NFW halo.  Both the
initial (solid curves) and remnant (dashed curves) galaxies are shown
and the mode effective radius $\re$ of each model's bulge is marked
with an asterisk.  The results for the initial models (solid lines)
show the dramatic effect of adiabatic contraction in the inner region
of a galaxy: the dark matter mass interior to $\re$ in the
adiabatically contracted model is approximately twice that of an
uncompressed NFW halo.  After the merger the dark matter fraction
interior to a given radius decreases, but the dark matter fraction
interior to the final $\re$ is \emph{greater} than the dark matter
fraction interior to the initial $\re$ for both simulations.  This
increasing dark matter fraction with stellar mass is consistent with
that needed to explain the observed tilt in the fundamental plane.
Quantitatively, we find $M_{dyn} \propto M_*^{1.11}$ for the
compressed haloes and $M_{dyn} \propto M_*^{1.19}$ for the
uncompressed haloes.  These results are reasonably consistent with the
observational results of \citet{pad04} from SDSS.  The amount of dark
matter within $\re$ can be constrained in other ways, particularly
through gravitational lensing.  Though a large, statistically
significant sample of lenses is needed, the dark matter fraction we
obtain is consistent with analyses of existing lensing observations
(e.g. \citealt{tk04,rk05}).

One way to understand the increasing dark matter fraction is based on
the dark matter mass profile in the inner parts of the galaxy: assume
that the dark matter mass profile is given by $M_{dm}(r) \propto M_v
r^\eta$.  For an NFW profile, $\eta \approx 2$ at the small radii
appropriate near $\re$.  For an adiabatically compressed halo, $\eta
\approx 1.2$ (Table \ref{table-rho}).  If the bulge stays roughly on
the $\re-M_*$ relation, $\re$ increases by a factor of $\approx 1.5$
during an equal-mass merger.  The stellar mass within $\re$ increases
by a factor of $2$ during such a merger, while the dark matter mass
within $\re$ increases by a factor of $\approx 2*1.5^\eta \approx
3.25-4.5$ if the virial mass also doubles.  If the final virial mass
is only a factor of $\approx 1.3$ of the initial virial mass (Table
\ref{table-rho}), the dark matter mass within $\re$ increases by a
smaller factor of $\approx 1.3*1.5^\eta \approx 2.1-3$.  All of
these estimates yield a dark matter fraction within $\re$ that
increases after the merger, consistent with our simulations.

\section{Summary \& Discussion}
\label{section:discuss}

We have performed numerical simulations of dissipationless mergers of
equal-mass galaxies containing a stellar bulge and a dark matter halo
to study how observable properties of elliptical galaxies scale during
mergers.  Some key features of our simulations include:

\begin{itemize}
  
\item We consider both standard NFW dark matter haloes and NFW haloes
  modified by adiabatic response to baryonic dissipation.
  
\item We set up the initial stellar bulges using the observed
  $\re-M_*$ projection of the fundamental plane (most of which also
  follow the observed $M_*-\sige$ relation).
  
\item Our primary merger orbit is taken from the most probable value
  of a cosmological distribution, which is weakly bound with
  $v_{\theta} \approx 0.7\,v_r$.  We also simulate a merger on a
  radial parabolic orbit and a merger on an orbit with the initial
  orbital velocities reduced by 15\% for comparison.
  
\item We perform a large number of random projections of each merger
  remnant in order to determine the effects of viewing angle on the
  observable characteristics.
  
\item Our resolution tests indicate that $\sim 5 \times 10^5$ dark
  matter particles per halo are necessary to accurately resolve the
  dynamics of the stellar component on the scale corresponding to
  observations.  Test runs with a smaller number of particles lead to
  artificially ``puffed up'' stellar remnants.

\end{itemize}

\noindent The main results of our work are:

\begin{itemize}
  
\item As in previous studies, we find that the remnant bulges retain a
  de Vaucouleurs profile. In addition, the spherically-averaged radial
  density profiles of the dark matter haloes appear to be scaled-up
  versions of those of their progenitors, i.e., with the dark matter
  density increasing by a simple normalization factor.

\item The remnant shapes depend strongly on the orbit of the merger.
  The cosmologically ``most probable'' orbit results in a nearly
  oblate halo with a triaxial bulge, while less probable radial
  orbits result in prolate bulges and haloes.
  
\item The $\re-M_*$ and $M_*-\sige$ projections of the \fpp\ are
  roughly preserved for an ``average'' viewing angle in major
  dissipationless mergers with the most probable orbit, independent of
  the global ratio of dark to luminous matter
  (Table~\ref{table-runs}).  There are, however, significant
  variations with viewing angle, and the observables are strongly
  correlated (Fig.~\ref{histmulti}).  The viewing-angle scatter
  appears consistent with the observed scatter in the \fpp.  Our
  simulations thus suggest that the \fpp\ is consistent with models in
  which massive elliptical galaxies undergo significant growth via
  late dissipationless mergers.  If the merging elliptical galaxies
  contain central massive black holes, and if black hole masses
  roughly add when they coalesce (i.e., there is not significant
  energy loss to gravitational waves), then our results also imply
  that dissipationless mergers maintain the $M_{\rm BH}-\sigma_e$
  (\citealt{geb00,fm00}) and $M_{\rm BH}-M_*$ (\citealt{mag98,hr04})
  relations.

\item The merger remnants contain more dark matter (relative to
  stellar matter) within $\re$ than their progenitors: $M_{dyn}
  \propto M_*^{1.11-1.19}$, consistent with the hypothesis that the
  tilt of the \fpp\ is due to an increasing dark matter fraction
  within $\re$.

\item An appreciable amount of energy and angular momentum are
  transfered from the stars to the dark matter in the merging process.
  Quantitatively, the energy transfer found in our simulations is
  similar to that needed in semi-analytic galaxy formation models to
  explain the \fpp.

\end{itemize}

One important goal of studies of elliptical galaxies is to understand
the origin of the tilt in the \fpp.  The virial theorem expectation
relates the stellar velocity dispersion and effective radius to the
\emph{total} mass within $\re$: $\sige^2 \propto M_{dyn}(<\re)/\re$.
The fundamental plane relates stellar velocity dispersion, luminosity,
and effective radius: $\re \propto \sige^{-3} L^{3/2}$.  As discussed
below equation~(\ref{MoverL}), if metallicity or stellar age
differences are solely responsible for the \fpp\ tilt then $M_{dyn}
\propto M_*$ and $M_* \propto L^x$ with $x \approx 1.15$.  On the
other hand, if dark matter fraction is responsible, then $M_* \propto
L$ and $M_{dyn} \propto M_*^x$ with $x \approx 1.15$.  Our simulations
contain no star formation or metallicity information and thus $M_*
\propto L$ by assumption; we therefore do not have a direct way to
evaluate the importance of metallicity variations (though we note that
there is observational evidence for them; e.g.
\citealt{trager,cap05}).  Nevertheless, we find that because of a
change in the amount of dark matter with $\re$ due to a merger,
dissipationless (major) mergers move elliptical galaxies roughly along
the \fpp.  It is interesting to note that the nearly homologous nature
of the bulges in our simulations indicates that our results may hold
for unequal mass mergers as well: using mass ratio $\xi=\frac{1}{3}$
rather than $\xi=1$ results in only a 1\% change in
$(f_{orb}+f_t)/f_i$ from equation~(\ref{eqn:homology}).

Because our simulations involve only gravitational interactions, the
results can be interpreted at different mass and length scales by an
appropriate rescaling of time.  This is somewhat complicated in
two-component simulations such as ours, however, because bulges and
dark matter haloes do not scale in the same manner: for example, $\re
\propto M_*^{0.56}$ while $r_v \propto M_v^{0.33}$.  If we rescale the
length and mass according to the virial theorem definition, then the
bulge no longer lies on the fundamental plane.  For example, consider
a mass scale 10 times larger than what is presented in this work
(keeping the concentration constant).  Both the virial radius and
effective radius increase by a factor of $10^{1/3}$; then for run M20,
the bulge effective radius will be $\re=10^{1/3} \times 2.81 \mbox{
  kpc}=6.05$ kpc rather than the 10.24 kpc predicted by the SDSS
scaling (equation~\ref{eqn:shen}).  This difference is
comparable to the $1-\sigma$ scatter Shen et al. find in the $\re-M_*$
relation.  As a result, we can indeed scale our simulations up or down
by a factor of 10 in mass and $10^{1/3}$ in radius and still be near
the fundamental plane.  Since the rescaling only affects the amplitude
of the relations, the slopes in equation~(\ref{eqn:fpslopes}) and
Table~\ref{table-runs} are actually preserved and the scaled mergers
would move along the fundamental plane but with a constant offset.

The results of Section \ref{shape} show that our remnant bulges are
either triaxial (non-radial orbit) or prolate (radial orbit), in
agreement with analyses of SDSS data.  In this context it is important
to point out that our initial conditions are spherical and isotropic,
while both the remnants of our mergers and observed ellipticals tend
to be triaxial.  It would certainly be desirable to use more realistic
(i.e. non-spherical) initial conditions in future work, though there
are significant difficulties in doing so: operationally, it is not
possible to create stable triaxial initial conditions corresponding to
a desired density profile.  It would also be of interest to extend the
results presented here by including a gaseous or stellar disk.  Such
simultions should help shed light on the origins of both spheriodal
systems and the fundamental plane, topics that we have not addressed
in this work.  We note that in simulations of mergers of
halo-bulge-disk systems performed by \citet{barnes92}, the stellar
remnant followed a similar radius-mass relation ($r \propto M^{0.5}$)
to what we find for halo-bulge mergers, perhaps implying that
inclusion of a collisionless stellar disk would not change the
conclusions of this work drastically, but this issue must be resolved
with future simulations.

\section*{Acknowledgments} 

We thank A. Dekel, S. Faber, L. Hernquist, N. Murray, \& T. Thompson
for useful discussions and Volker Springel for making GADGET
available.  This research used resources of the National Energy
Research Scientific Computing Center, which is supported by the Office
of Science of the U.S. Department of Energy.  CPM is partially
supported by NSF grant AST 0407351 and NASA grant NAG5-12173.  EQ is
supported in part by NSF grant AST 0206006, NASA grant NAG5-12043, an
Alfred P. Sloan Fellowship, and the David and Lucile Packard
Foundation.

\appendix
\section{Sersic fitting}
\label{section:re}
\begin{figure}
\begin{center}\leavevmode
\includegraphics[scale=0.8]{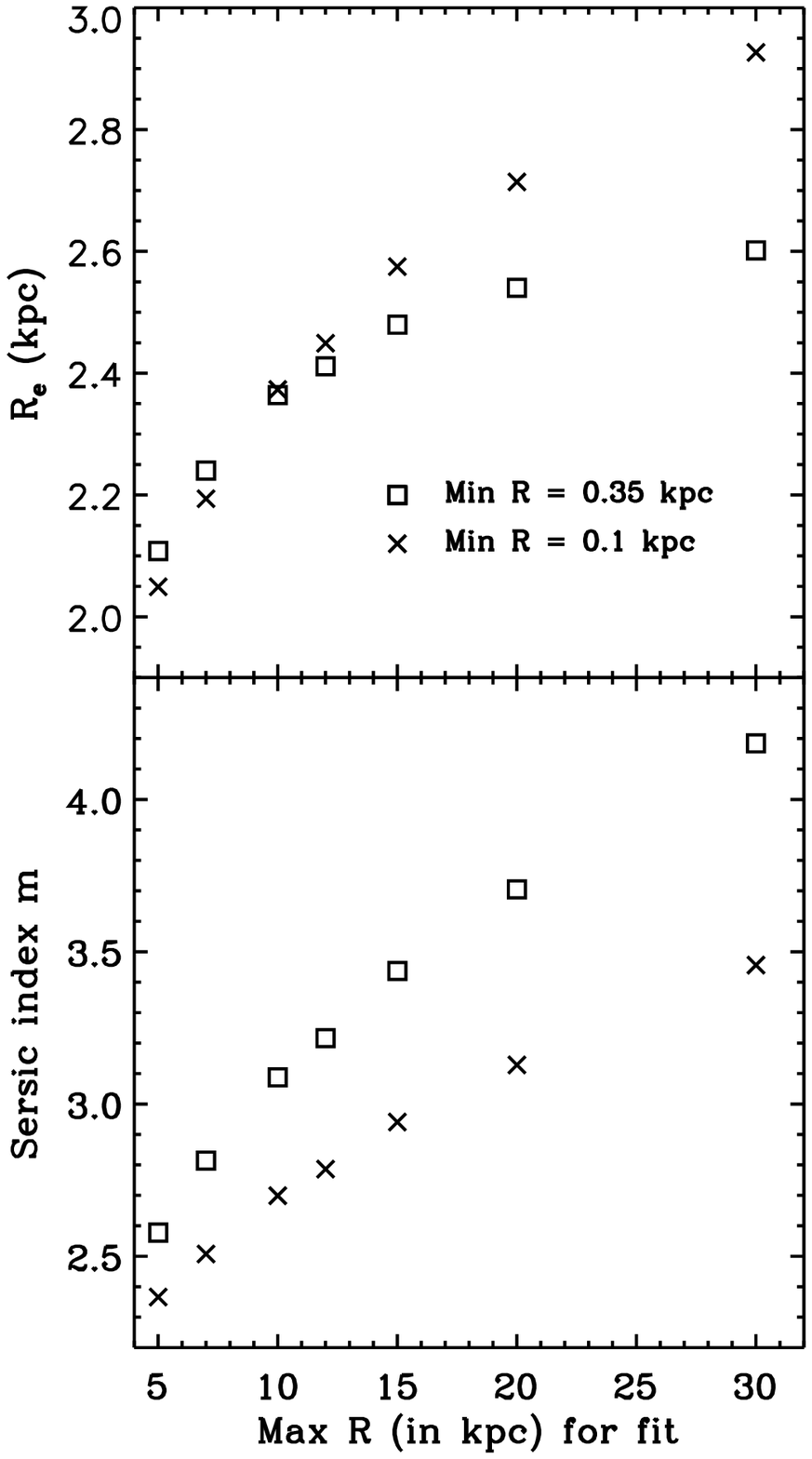}
\end{center}
\caption{A demonstration of the dependence of $\re$ and $m$ obtained from
  Sersic fitting (equation~\ref{sers}) on the radial range of the fit for
  a Hernquist profile with $\re=2.81$ kpc.  The top panel plots the
  derived effective radius $\re$ as a function of the maximum radius
  to which the fit is performed while the bottom panel plots the
  fitted Sersic index $m$.  The best-fit values of both $R_e$ and $m$
  depend sensitively on both the maximum (horizontal axis) and minimum
  (symbols) radii used in the fit.}
\label{sersic}
\end{figure}
A common parameterization of the surface brightness profile of a galaxy
is that of \citet{sersic}:
\begin{equation}
    I(R) = I(\re) \exp\{-b(m)[(R/\re)^{1/m}-1]\}. 
\label{sers}
\end{equation}
Though the function $b(m)$ cannot be expressed in closed form, an
accurate asymptotic expansion is given by \citet{cb99} as $b(m)
\approx 2m-1/3+4/(405m)$; bulges and early-type galaxies are typically
well-approximated by the \citet{dv48} law of $m=4$.  Fitting an
observed surface brightness distribution to a Sersic profile is a
standard technique for determining the (projected) half-mass radius
$\re$ of a galaxy.  We find that fitting our projected density
profiles to Sersic profiles is not robust, however: changing the
radial range of the fit leads to systematic variations in both the
Sersic index and the corresponding effective radius.

This dependence of fitting results on the range of radii can be seen
for the initial bulge that is assigned a Hernquist profile.
Fig.~\ref{sersic} shows the derived $\re$ (top panel) and $m$ (bottom
panel) for an input Hernquist profile with $\re=2.81$ kpc as a
function of the maximum radius of the fit.  Two different values for
the minimum radius of the fit are plotted: 0.1 kpc (X symbols) and
0.35 kpc (square symbols).  In both cases the recovered $\re$ and $m$
are increasing functions of the maximum radius of the fit;
additionally, the recovered $\re$ is lower than the input $\re=2.81$
kpc and the Sersic index $m$ is less than 4 for all reasonable maximum
values of the fit.  This result is not unexpected since Hernquist
models deviate from the de Vaucouleurs profile at both small and large
radii, but it does present a problem for quantitatively characterizing
the stellar bulges in our simulations since the fitted $\re$ for the
initial Hernquist profile can change by 20\% or more depending on the
range of radii considered; the final bulges show similar trends.
In fact, \citet{blanton05} cite a similar issue with Sersic fits to
SDSS observations (see their appendix A and fig. 10).

As a result, we do not use Sersic fits to determine the effective
radii of either our initial conditions or merged ellipsoids.  Instead,
we take a line of sight and calculate the surface mass density
perpendicular to the line of sight in spherical bins.  We also
calculate the projected enclosed mass at each radius.  We define the
projected radius of the particle enclosing half the total mass to be
the half-light radius ($\re$).  This procedure recovers the correct
$\re$ for the initial hernquist profile to within 1-2~\%.

\section{Energy Conservation}
\label{appendix:energy}
Consider a system of particles with distribution function $f$ obeying
the collisionless Boltzmann equation
\begin{equation}
\label{eqn:CBE}
\frac{\partial f}{\partial t}+\vec{v} \cdot \frac{\partial f}
{\partial \vec{x}} - \frac{\partial \Phi} {\partial \vec{x}}\cdot
\frac{\partial f} {\partial \vec{v}}=0.
\end{equation}
By taking, in turn, a velocity and spatial moment of
equation~(\ref{eqn:CBE}) 
we can obtain the tensor virial theorem
(\citealt{bt87}, sec. 4.3):
\begin{equation}
\label{eqn:virial}
\frac{1}{2} \frac{d^2I_{\alpha \beta}}{dt^2}=2K_{\alpha \beta}+
W_{\alpha \beta},
\end{equation}
where 
\begin{eqnarray}
I_{\alpha \beta} &=& \int \rho x_\alpha x_\beta d^3x,\\
K_{\alpha \beta} &=& \frac{1}{2} \int \rho \overline{v_\alpha v_\beta}
d^3x,\\
W_{\alpha \beta} &=& -\int \rho x_\beta \frac{\partial \Phi} {\partial
x_\alpha}d^3x 
\end{eqnarray}
and $\overline{v_\alpha v_\beta}$ is the mean of $v_\alpha v_\beta$.
For a steady-state system, the second derivative of the inertia tensor
$I_{\alpha \beta}$ vanishes and the trace of
equation~(\ref{eqn:virial}) gives the standard scalar virial theorem
$2K=-W$.  If we consider a system with a bulge (b) and dark matter
(d), each with a distribution function obeying the collisionless
Boltzmann equation with respect to the combined
potential\footnote{Since phase space densities are additive, the total
  phase space density $f_{tot}=f_b+f_d$ also obeys the collisionless
  Boltzmann equation.}, we can write the same tensor virial theorem
for each component by replacing the total mass density $\rho$ with the
mass density of that component.  This modification is straightforward
for the inertia tensor and kinetic term but is somewhat more subtle
for the potential term $W_{\alpha \beta}$.  By writing the potential
as the sum of the potential of each subsystem, $\Phi=\Phi_b+\Phi_d$,
the trace for the bulge breaks into two separate parts:
\begin{equation}
W_b=-\int \rho_b \vec{x} \cdot \frac{\partial \Phi_b}{\partial \vec{x}}d^3{x}
-\int \rho_b \vec{x} \cdot \frac{\partial
\Phi_d}{\partial\vec{x}}d^3{x}.
\end{equation}
The first term in the above equation, which we denote $W_{bb}$, is the
familiar self-energy of the bulge.  The second term, $W_{bd}$,
describes the interaction energy of the bulge with the dark matter
halo.  If we write the mass density as a sum over (Dirac) delta
functions
\begin{equation}
\rho_b=\sum_{i \in b} m_i \delta(\vec{x} -\vec{x}_i)
\end{equation}
and similarly for component $d$, then this term takes a more
transparent form:
\begin{equation}
\label{eqn:pot_sum}
W_{bd}=-\sum_{i \in b} \sum_{j \in d} G m_i m_j~\vec{x}_i \cdot
\frac{(\vec{x}_i - \vec{x}_j)}{|\vec{x}_i - \vec{x_j}|^3}.
\end{equation}

This quantity $W_{bd}$ is energy belonging to the bulge due to the
interaction of the bulge with the dark matter such that the total
potential energy of the bulge is defined as $W_b \equiv
W_{bb}+W_{bd}$.  [The second term in equation~(\ref{eqn:Econs}) is $W_{bb}$
while the third term is $W_{bd}$.]  We similarly define the total
potential energy of the dark matter halo to be $W_d \equiv W_{dd} +
W_{db}.$ Note with this definition we have
\begin{equation}
W_{bd}+W_{db}=-\sum_{i \in b} \sum_{j \in d} G m_i m_j
\frac{1}{|\vec{x}_i - \vec{x}_j|}, 
\end{equation} 
which is simply the ``cross'' potential energy of the
bulge and dark matter components, and thus
\begin{equation}
W_b+W_d=W_{bb}+W_{bd}+W_{dd}+W_{db}=W_{\mathrm{tot}}
\end{equation}
where $W_{\mathrm{tot}}$ is the total potential energy of the system.
For each galaxy used in our initial conditions,
$2~T_{bb}/|W_{bb}+W_{bd}|=1$ to within 1\%.
\label{lastpage}


\begin{thebibliography}{}
\bibitem[\protect\citeauthoryear{Alam \& Ryden}{2002}]{alam02}
Alam S. M. K., Ryden B. S. 2002, ApJ, 570, 610

\bibitem[\protect\citeauthoryear{Bailin \& Steinmetz}{2004}]{bs04}
Bailin J., Steinmetz M., 2005, ApJ submitted (astro-ph/0408163)

\bibitem[\protect\citeauthoryear{Barnes}{1992}]{barnes92}
Barnes J. E. 1992, ApJ, 393, 484

\bibitem[\protect\citeauthoryear{Barnes \& Hernquist}{1992}]{bh92}
Barnes J. E., Hernquist L. 1992, ARA\&A, 30, 705

\bibitem[\protect\citeauthoryear{Benson}{2005}]{benson04}
Benson A. J. 2005, MNRAS, 358, 551

\bibitem[\protect\citeauthoryear{Bernardi et al.}{2003a}]{bernardi2}
Bernardi M. et al., 2003a, AJ, 125, 1849

\bibitem[\protect\citeauthoryear{Bernardi et al.}{2003b}]{bernardi3}
Bernardi M. et al., 2003b, AJ, 125, 1866

\bibitem[\protect\citeauthoryear{Binney \& Tremaine}{1987}]{bt87}
Binney J., Tremaine S. 1987, Galactic Dynamics, Princeton
University Press, Princeton, N. J.

\bibitem[\protect\citeauthoryear{Blanton et al.}{2005}]{blanton05}
Blanton M. et al 2005, ApJ in press (astro-ph/0310453)

\bibitem[\protect\citeauthoryear{Blumenthal et al.}{1986}]{blum86}
Blumenthal G. R., Faber S. M., Flores R., Primack J. R. 1986, ApJ,
301, 27

\bibitem[\protect\citeauthoryear{Borriello, Salucci, \& Danese}{2003}]
  {borriello03} 
Borriello A., Salucci P., Danese L. 2003, MNRAS, 341, 1109

\bibitem[\protect\citeauthoryear{Boylan-Kolchin \& Ma}{2004}]{mrbk04}
Boylan-Kolchin M., Ma C.-P., 2004, MNRAS, 349, 1117

\bibitem[\protect\citeauthoryear{Burkert \& Naab}{2003}]{bn03} 
  Burkert A., Naab T. 2003, in `"Galaxies and Chaos", eds. G.
  Contopoulos and N. Voglis (Springer); astro-ph/0301385

\bibitem[\protect\citeauthoryear{Capelato, de Carvalho, \&
Carlberg}{1995}]{ccc95}
Capelato H. V., de Carvalho R. R., Carlberg R. G. 1995, ApJ, 451, 525

\bibitem[\protect\citeauthoryear{Cappellari et al.}{2005}]{cap05}
 Cappellari M. et al., MNRAS submitted (astro-ph/0505042)

\bibitem[\protect\citeauthoryear{Ciotti \& Bertin}{1999}]{cb99}
Ciotti L, Bertin G. 1999, A \& A, 352, 447

\bibitem[\protect\citeauthoryear{Cole et al.}{2000}]{cole00}
Cole S., Lacey C. G., Baugh C. M., Frenk C. S. 2000, MNRAS, 319, 168

\bibitem[\protect\citeauthoryear{Dantas et al.}{2003}]{dantas03}
Dantas C. C., Capelato H. V., Ribeiro A. L. B., de Carvalho
R. R. 2003, MNRAS 340, 398

\bibitem[\protect\citeauthoryear{de Vaucouleurs}{1948}]{dv48}
de Vaucouleurs G. 1948, Ann. d'Astrophys., 11, 247

\bibitem[\protect\citeauthoryear{Dehnen}{1993}]{dehnen93}
Dehnen W. 1993, MNRAS, 265, 250 


\bibitem[\protect\citeauthoryear{Djorgovski \& Davis}{1987}]{dd87}
Djorgovski S., Davis M. 1987, ApJ, 313, 59

\bibitem[\protect\citeauthoryear{Dressler et al.}{1987}]{dressler87}
Dressler A., Lynden-Bell D., Burstein D., Davies R. L., Faber S. M.,
  Terlevich R. J., Wegner G. 1987, ApJ, 313, 42

\bibitem[\protect\citeauthoryear{Dubinski}{1998}]{dubinski98}
Dubinski J. 1998, ApJ, 502, 141

\bibitem[\protect\citeauthoryear{Dubinski \& Carlberg}{1991}]{dc91}
Dubinski J., Carlberg R. G. 1991, ApJ, 378, 496

\bibitem[\protect\citeauthoryear{Faber \& Jackson}{1976}]{fj76}
Faber S. M., Jackson R. E., 1976, ApJ, 204, 668

\bibitem[\protect\citeauthoryear{Faber et al.}{1997}]{faber97}
Faber S. M., Tremaine S., Ajhar E. A., Byun Y. I., Dressler A.,
Gebhardt K., Grillmair C., Kormendy J., Lauer T. R.,  Richstone
D. 1997, AJ 114, 1771 

\bibitem[\protect\citeauthoryear{Ferrarese \& Merritt}{2000}]{fm00}
Ferrarese L., Merritt, D. 2000, ApJ, 539, L9

\bibitem[\protect\citeauthoryear{Gao et al.}{2004}]{gao04}
Gao L., Loeb A., Peebles P. J. E., White S. D. M, Jenkins A. 2004,
ApJ, 614, 17

\bibitem[\protect\citeauthoryear{Gebhardt et al.}{2000}]{geb00}
Gebhardt K. et al., ApJ, 539, L13

\bibitem[\protect\citeauthoryear{Gerhard et al.}{2001}]{gerhard01}
Gerhard O., Kronawitter A., Saglia R. P., Bender R. 2001, AJ, 121, 1936

\bibitem[\protect\citeauthoryear{Gnedin et al.}{2004}]{gkkn04}
Gnedin O. Y., Kravtsov A. V., Klypin A. A., Nagai D. 2004, ApJ, 616, 16

\bibitem[\protect\citeauthoryear{Gonz\'{a}lez-Garc\'{i}a \& van
Albada}{2003}]{ggva}
Gonz\'{a}lez-Garc\'{i}a A. C., van Albada T. S. 2003, MNRAS, 342, L36

\bibitem[\protect\citeauthoryear{Graham \& Colless}{1997}]{gc97}
Graham A., Colless M. 1997, MNRAS, 287, 221

\bibitem[\protect\citeauthoryear{H\"{a}ring \& Rix}{2004}]{hr04}
H\"{a}ring N., Rix H.-W. 2004, ApJ, 604, L89

\bibitem[\protect\citeauthoryear{Hernquist}{1990}]{hern90} 
Hernquist L.  1990, ApJ, 356, 359


\bibitem[\protect\citeauthoryear{Jing \& Suto}{2002}]{js02}
Jing Y. P. \& Suto Y. 2002, ApJ, 574, 538

\bibitem[\protect\citeauthoryear{Jorgensen, Franx, \&
    Kjaergaard}{1996}] {jorgensen96} Jorgensen I., Franx M.,
  Kjaergaard P. 1996, MNRAS, 280, 167

\bibitem[\protect\citeauthoryear{Kauffmann et al.}{2003}]{kauffmann03}
Kauffmann G. et al. 2003, MNRAS 341, 33

\bibitem[\protect\citeauthoryear{Kazantzidis et al.}{2004}]{kazan04}
Kazantzidis S., Kravtsov A. V., Zentner A. R., Allgood B., Nagai D.,
\& Moore B. 2004, ApJ, 611, L73

\bibitem[\protect\citeauthoryear{Kazantzidis, Magorrian, \& Moore}{2004}]
{kmm04}
Kazantzidis S., Magorrian J., Moore B. 2004, ApJ 601, 37

\bibitem[\protect\citeauthoryear{Khochfar \& Burkert}{2003}]{kb03}
Khochfar S., Burkert A. 2003, ApJ, 597, L117

\bibitem[\protect\citeauthoryear{Khochfar \& Burkert}{2005}]{kb05}
Khochfar S., Burkert A. 2005, submitted to MNRAS (astro-ph/0309611)

\bibitem[\protect\citeauthoryear{Magorrian et al.}{1998}]{mag98}
Magorrian J. et al. 1998, AJ, 115, 2285

\bibitem[\protect\citeauthoryear{Moore et al.}{2004}]{moore04}
Moore B., Kazantzidis S., Diemand J., Stadel J. 2004, MNRAS, 354, 522

\bibitem[\protect\citeauthoryear{Naab \& Burkert}{2003}]{nb03}
Naab T., Burkert A. 2003, ApJ, 597, 893

\bibitem[\protect\citeauthoryear{Navarro, Frenk, \& White}{1997}]{nfw97}
Navarro J. F., Frenk C. S., White S. D. M. 1997, ApJ, 490, 493

\bibitem[\protect\citeauthoryear{Nipoti, Londrillo, \& Ciotti}{2003}]
{nipoti03}
Nipoti C., Londrillo P., Ciotti L. 2003, MNRAS, 342, 501

\bibitem[\protect\citeauthoryear{Padmanabhan et al.}{2004}]{pad04}
Padmanabhan N. et al. 2004, New Astronomy, 9, 329

\bibitem[\protect\citeauthoryear{Pahre, de Carvalho, \&
Djorgovski}{1998}]{pahre98a}
Pahre M. A., de Carvalho R. R., Djorgovski S. G. 1998, AJ, 116, 1591


\bibitem[\protect\citeauthoryear{Pahre, de Carvalho, \&
Djorgovski}{1998}]{pahre98b}
Pahre M. A., de Carvalho R. R., Djorgovski S. G. 1998, AJ, 116, 1606

\bibitem[\protect\citeauthoryear{Rusin \& Kochanek}{2005}]{rk05}
Rusin D., Kochanek C. S. 2005, ApJ, 623, 666

\bibitem[\protect\citeauthoryear{Seljak}{2002}]{seljak02}
Seljak U. 2002, MNRAS, 334, 797

\bibitem[\protect\citeauthoryear{Sersic}{1968}]{sersic} 
Sersic J. L.,1968, Atlas de Galaxies Australes. Observatorio
Astronomico, Cordoba 

\bibitem[\protect\citeauthoryear{Shen et al.}{2003}]{shen03}
Shen S., Mo H. J., White S. D. M., Blanton M. R., Kauffmann G., Voges
W., Brinkmann J., Csabai I. 2003, MNRAS, 343, 978

\bibitem[\protect\citeauthoryear{Springel \& Hernquist}{2005}]{sh05}
Springel V., Hernquist L. 2005, ApJ, 622, L9

\bibitem[\protect\citeauthoryear{Springel \& White}{1999}]{sw99}
Springel V., White S. D. M. 1999, MNRAS, 307, 162

\bibitem[\protect\citeauthoryear{Springel, Yoshida, \& White}{2001}]{gadget}
Springel V., Yoshida N., White S. D. M. 2001, New Astronomy, 6, 79

\bibitem[\protect\citeauthoryear{Steinmetz \& Navarro}{2002}]{sn02}
Steinmetz M., Navarro J. F. 2002, New Astronomy, 7, 155

\bibitem[\protect\citeauthoryear{Tormen}{1997}]{tormen97}
Tormen G. 1997, MNRAS, 290, 411

\bibitem[\protect\citeauthoryear{Trager et al.}{2000}]{trager}
Trager S. C., Faber S. M,  Worthey G., Gonz\'{a}lez J. 2000, AJ, 120, 165

\bibitem[\protect\citeauthoryear{Tremaine et al.}{1994}]{tremaine94}
Tremaine S., Richstone D. O., Byun Y., Dressler A.,  Faber S. M.,
Grillmair C., Kormendy J., Lauer T. R., 1994, AJ, 107, 634

\bibitem[\protect\citeauthoryear{Treu \& Koopmans}{2004}]{tk04}
Treu T., Koopmans L. V. E. 2004, ApJ, 611, 739

\bibitem[\protect\citeauthoryear{Toomre \& Toomre}{1972}]{toomre}
Toomre A., Toomre J. 1972, ApJ, 178, 623 

\bibitem[\protect\citeauthoryear{van Dokkum et al.}{1999}]{vd99}
van Dokkum P. G., Franx M., Fabricant D., Kelson D. D., Illingworth
G. D., 1999, ApJ, 520, L95

\bibitem[\protect\citeauthoryear{Vincent \& Ryden}{2005}]{vr05}
Vincent R. A., Ryden B. S. 2005, ApJ, 623, 137 

\bibitem[\protect\citeauthoryear{Zentner et al.}{2005}]{zentner05}
Zentner A., Berlind A. A., Bullock J. S., Kravtsov A. V., Wechsler
R. H. 2005, ApJ, 624, 505 

\end{thebibliography}
\end{document}